\documentclass[letterpaper,10pt]{article} 

\usepackage{opticameet3} 

\newcommand\authormark[1]{\textsuperscript{#1}}

\usepackage{amsmath,amssymb}
\usepackage[colorlinks=true,bookmarks=false,citecolor=blue,urlcolor=blue]{hyperref} 

\begin{document}

\title{High Resolution On-Chip Thin-Film Lithium Niobate Single-Photon Buffer}


\author{Cagin Ekici,\authormark{1} Yonghe Yu,\authormark{1} Jeremy C. Adcock,\authormark{1} Alif Laila Muthali,\authormark{1} Heyun Tan,\authormark{2}  Hao Li,\authormark{2}  Leif Katsuo Oxenløwe,\authormark{1} Xinlun Cai,\authormark{2} and Yunhong Ding\authormark{1,*}}

\address{\authormark{1} Center for Silicon Photonics for Optical Communication (SPOC), Department of Electrical and Photonics Engineering, Technical University of Denmark, Lyngby, Denmark\\
\authormark{2}State Key Laboratory of Optoelectronic Materials and Technologies, School of Electronics and Information Technology, Sun Yat-sen University,
Guangzhou 510275, China}

\email{\authormark{*}yudin@dtu.dk} 

\begin{abstract}
 We experimentally demonstrate a room-temperature, voltage controlled, short-term quantum photonics memory on a lithium niobate chip. Our chip is capable of resolving 100 ps time steps with 0.74 dB loss per round-trip.
\end{abstract}

\section{Introduction}
Short-term quantum photonics memories or single-photon buffers are essential for quantum technologies, since they provide a synchronization scheme for matching independent systems functioning at different speeds. In order to optimize two-photon interference from distant sources in a quantum network, photon buffers having high resolution configurability is needed to store one photon until the other is transmitted \cite{Azuma15}. In addition, entangling quantum operations in photonics are generally probabilistic, and such short-term memories play a crucial role to buffer gates of probabilistic nature. Furthermore, approaching ideal single-photon sources based on parametric spontaneous pair generation through temporal multiplexing requires low-loss and controllable photon storage \cite{Adcock22,AdcockQST22}.

 To date, optical buffers based on delay lines \cite{Burmeister08}, slow light \cite{Tucker05}, and Bragg scattering four-wave mixing \cite{Clemmen18} have been introduced. All these techniques either have an excessive loss which is not suitable for quantum applications or are overly sophisticated. Although atomic cloud optical memories are main contenders, they are difficult to integrate, and only operate at specific wavelengths. Therefore, to fulfill the requirements of a single-photon buffer, thin-film lithium niobate (TFLN) based integrated photonics platforms are ideal candidates, since they offer voltage controlled, low-loss and high-speed interferometric switching. In this paper, we experimentally demonstrate an on-chip TFLN single-photon buffer based on recirculating 1 cm-long loop with a round-trip time of 100 ps, i.e. the overall delay can be controlled with 100 ps time resolution, and storage times of up to 1.4 ns (14 round trips).

\section{Experimental Setup and Results}
\begin{figure}[!htbp]
\centering
\mbox{%
\includegraphics[scale=0.058]{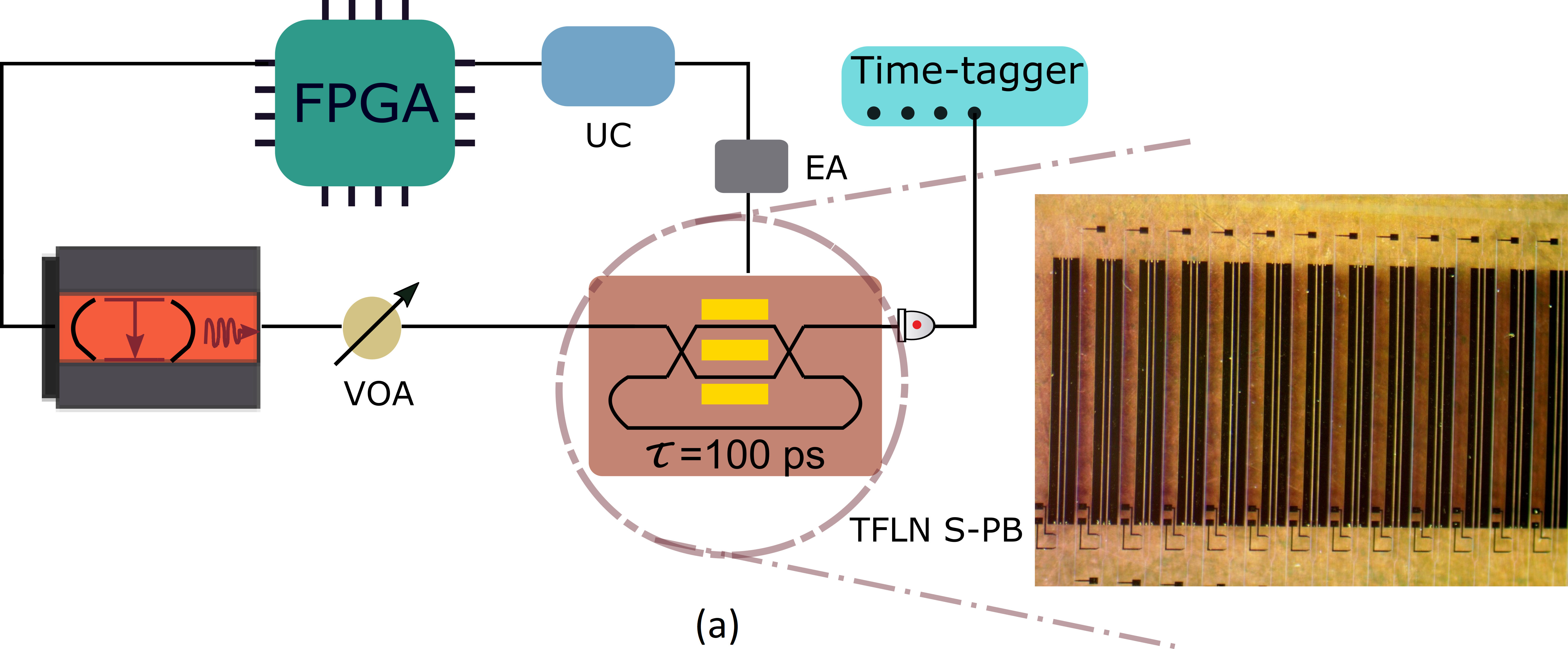}\hskip1em\includegraphics[scale=0.26]{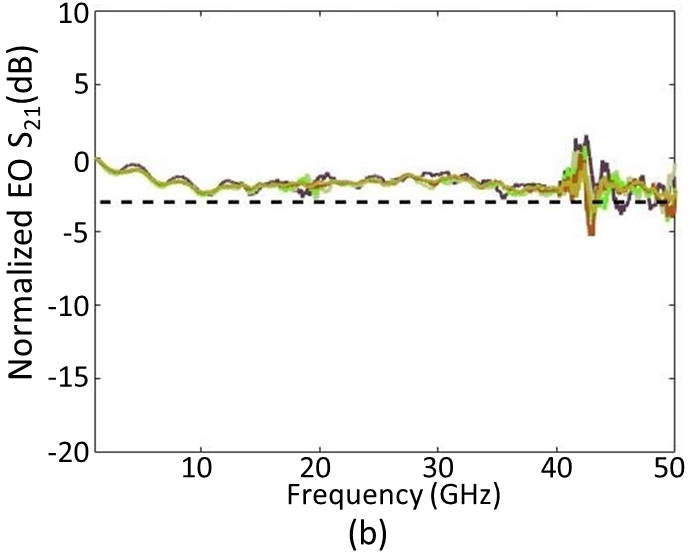}
}

\caption{\label{exp} (a) Schematics of the experimental setup with a real image of the TFLN chip consisting several buffers. (b) Electro-optic bandwidth (S$_{21}$) measurement.  Abbreviations: FPGA: Field-Programmable Gate Array, VOA: Variable Optical Attenuator, UC: Ultrafast Comparator, EA: Electronic Amplifier, TFLN S-PB: TFLN Single-Photon Buffer.}
\end{figure}

The TFLN single-photon buffer was fabricated on a commercial lithium niobate on insulator (LNOI) platform with top LN thickness of 600 nm. The switch consists of 4.5~mm long LN phase modulator on push-pull mode, exhibiting bandwidth more than 40~GHz, as shown in Fig. \ref{exp} (b), and the whole chip insertion loss is less than 6.2 dB (including the coupling loss). 

The experimental setup is shown in Fig. \ref{exp} (a). We conduct the experiments utilizing heavily-attenuated light from a laser (1550 nm, 40 fs pulse duration), i.e. weak coherent state, with 100 MHz repetition rate instead of true single-photon quantum states. The switch control signals are generated via an FPGA and are fed into an ultrafast comparator to obtain a fast fall-rise time. Afterwards, the fast signals are amplified to the $V_{\pi}$ of the TFLN switch, and are applied to the chip through high speed radio frequency (RF) probes using micropositioners. After storage and read-out for a delay, photons are detected by a superconducting nanowire single-photon detector(s) and recorded by a time-tagger which produces a real-time histogram of the detection event. 


\begin{figure}[!htbp]
\centering
\mbox{%
\includegraphics[scale=0.079]{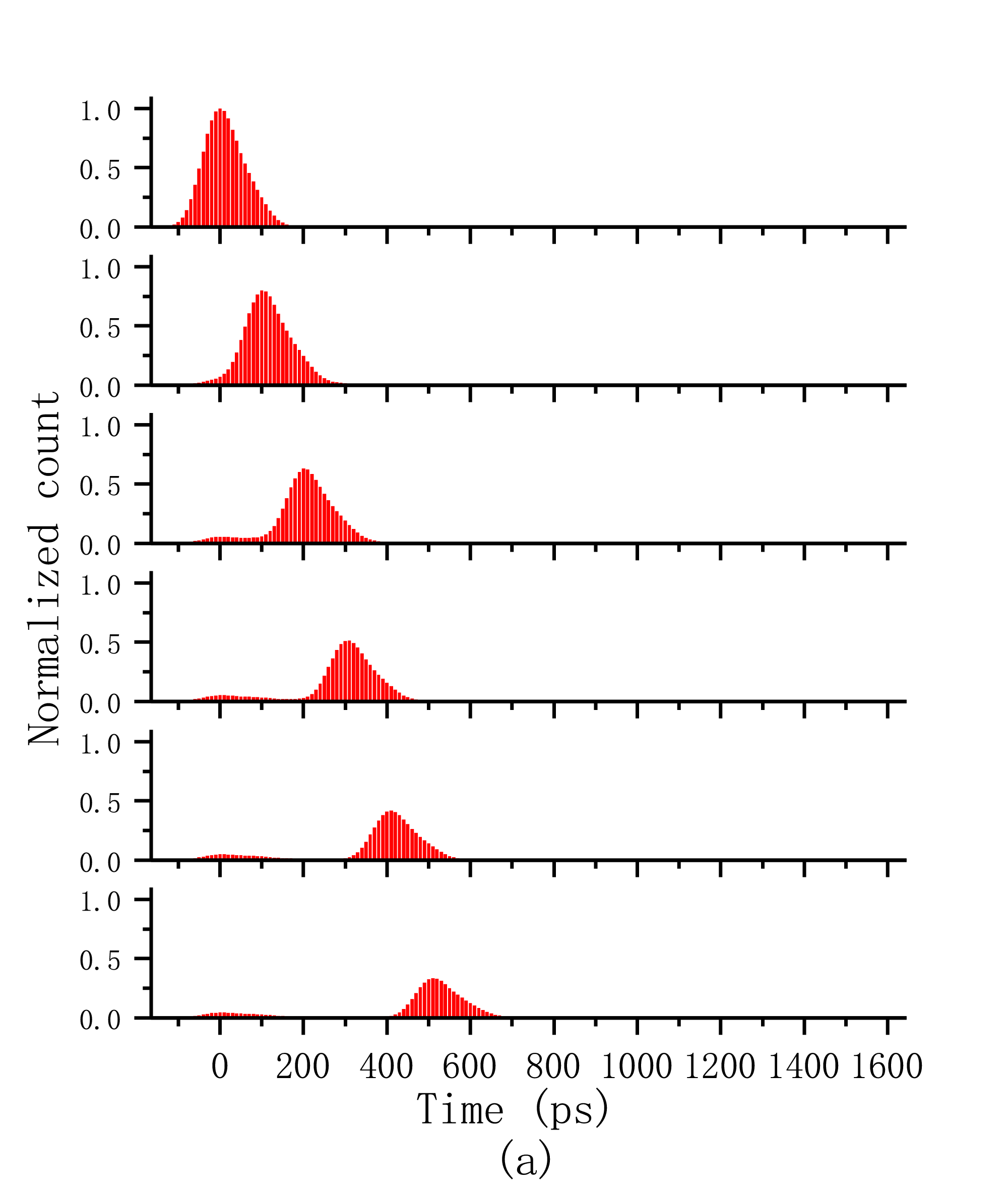}\hskip-1em\includegraphics[scale=0.079]{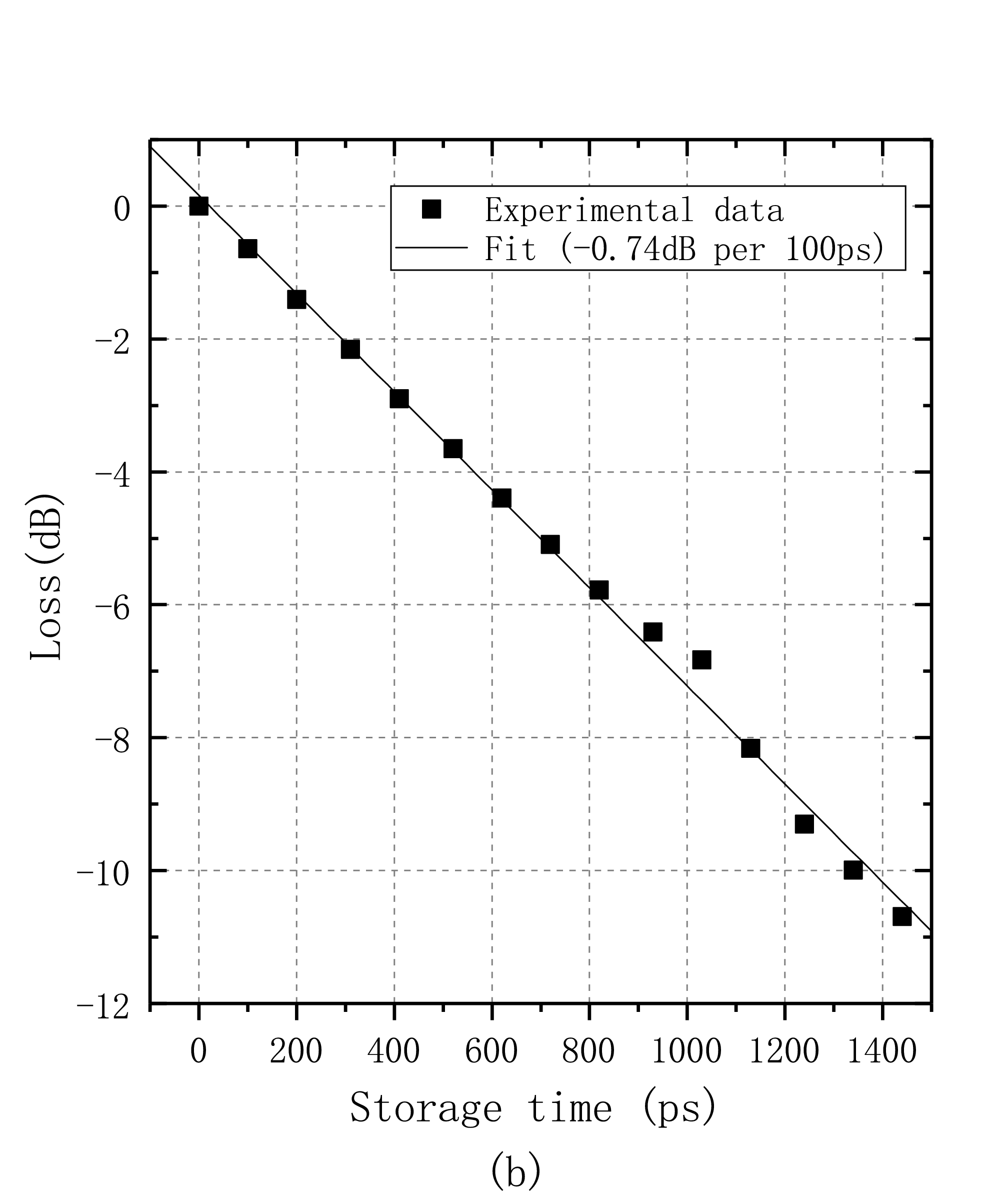}\hskip-1em\includegraphics[scale=0.079]{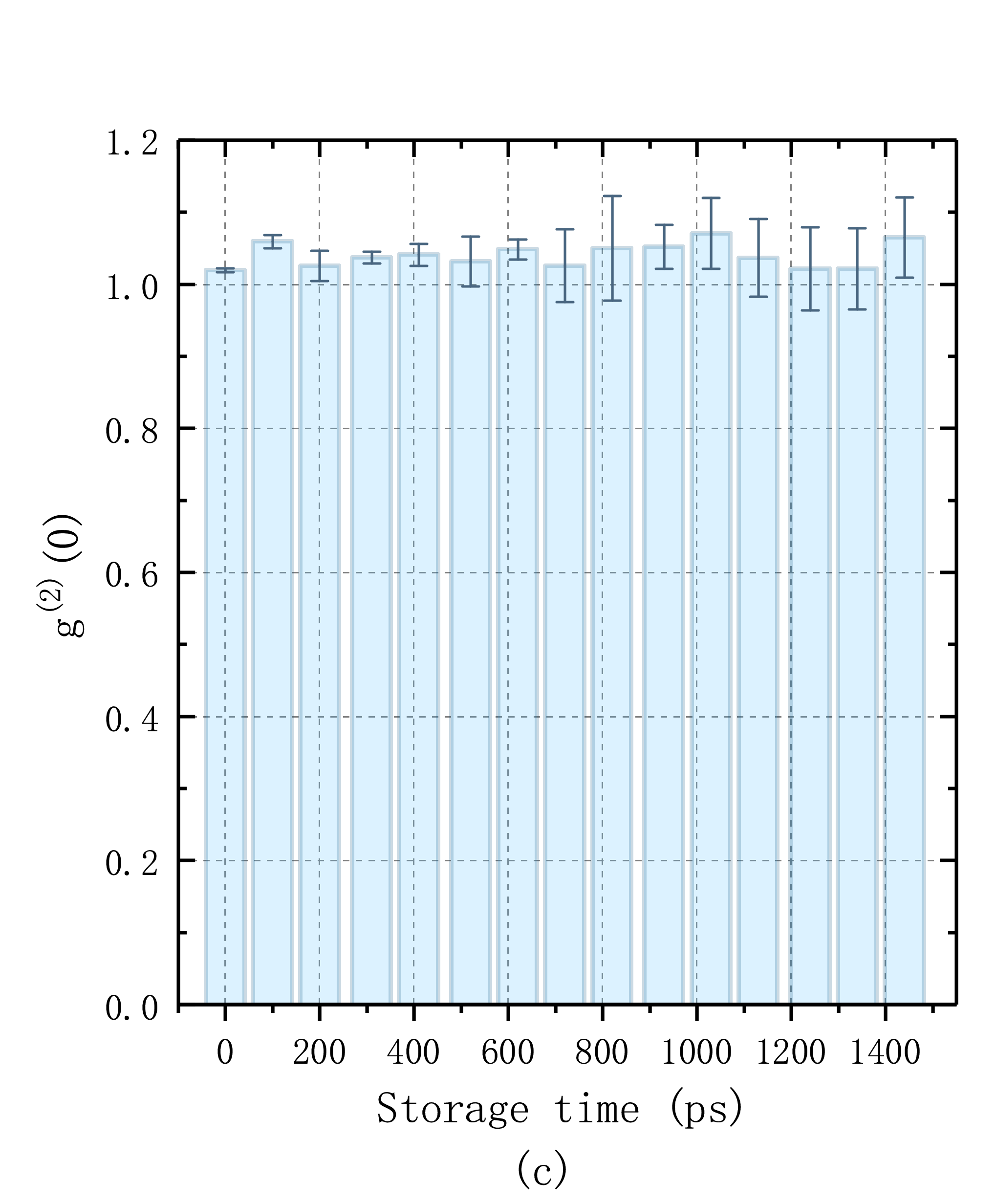}
}

\caption{\label{lossg2} The experimental results of single-photon storage: (a) Normalized histogram counts as a function of different storage time. (b) The peak amplitudes of the normalized histogram counts revealing the loss for different storage times. (c) Second-order correlation function $g^{(2)}(0)$ of the read-out single photons for different storage times with the error bars.}
\end{figure}

The experimental results of single photon storage with our TFLN chip is shown in Fig. \ref{lossg2}. Normalized histogram counts for the first 5 round-trip are depicted in Fig. \ref{lossg2} (a). The round-trip loss performance of the chip as a function of time is exhibited in Fig. \ref{lossg2} (b). Each peak value after a round-trip has been fitted the line with slope 0.74 dB.  Accordingly, we measure the second-order correlation function $g^{(2)}(0)$ after each round-trip by adding a 50/50 fiber optic beam splitter before the detection, see Fig. \ref{lossg2} (c). As expected $g^{(2)}(0)\approx 1$, since our TFLN photonics chip is illuminated by a weak coherent state. As a result of constant $g^{(2)}(0) \approx 1$ for every round-trip, it can be inferred that the statistics do not change significantly as a function of a storage time and there is no substantial optical background noise owing to the absence of an optical pump beam \cite{Kupchak15}.

\section{Conclusion}
We present an experimental study of a recirculating on-chip TFLN single-photon buffer enabling single photons to be captured, stored, and read-out at will with 100 ps time step resolution in a reliable way. Our promising chip is a robust and scalable architecture working at room-temperature with low-loss around 0.74 dB per round-trip.

\end{document}